\documentclass{article}
\usepackage{amssymb,latexsym, amscd}

\begin{document}

\title{Group contractions and its consequences upon representations of different
spatial symmetry groups.}

\author{Mauricio Ayala-S\'anchez$^{a}$\footnote{Email: \texttt{mauricio@ayala.as}}
\\ and 
\\ Richard W. Haase$^{b,c}$\footnote{Email: \texttt{rhaase@ciencias.unal.edu.co}}
\\
\\$^{a}$ Departamento de Matem\'aticas, 
\\ Universidad de Los Andes, Bogot\'a - Colombia.
\\
\\$^{b}$ Departamento de F\'{\i}sica, 
\\ Universidad Nacional de Colombia, Bogot\'a - Colombia.
\\
\\$^{c}$ Centro Internacional de F\'{\i}sica, A.A.4948, Bogot\'a - Colombia.
}

\date{}

\maketitle

\abstract{
We investigate the group contraction method for various space-time groups, including [$SO_3\to
E_2$], [$SO_{3,1}\to G_3$], [$SO_{5-h,h}\to P_{3,1}$ ($h=1,2$)], and its consequences for
representations of these groups. Following strictly quantum mechanical procedures we
specifically pay attention in the asymptotic limiting procedure employed in the contraction
[$G\to G'$], not only to the respective algebras but to their representations spaces spanned by
the eigenvectors of the Cartan subalgebra and the eigenvalues labelling these representation
spaces. Where appropriate a physical interpretation is given to the contraction prodecure.}

\section{Introduction}

The group contraction method was initially introduced by In\"on\"u and Wigner in 1953
\cite{Inonu-Wigner}, \cite{Inonu}, \cite{Gilmore}. The main interest resulted from the study of the transition
of relativistic to nonrelativistic quantum mechanics in the asymptotic limit when velocities are
small compared to the velocity of light. Under this limit the Lorentz group becomes the Homogeneous Galilei group. The group contraction method continues to be a subject of active research in particularly as applied to quantum groups. In general this method determines under a change of parameter scale a limiting process on the group generators and its algebra producing as a consequence a new group with corresponding algebra. The main difficultly in a general approach is the construction of the sequence of limiting  representation spaces. In this paper we focus on this aspect of the problem by reviewing some well-know applications to space-time symmetry groups. We employ a direct approach by which we apply the limiting procedure to the standard quantum mechanical treatment of the construction of representation spaces, ie. as an eigensystem problem of the generators of the algebra, {\it \'a la} $SO_3$ of angular momentum theory.

Since a Lie group is uniquely determined by its algebra about the identity, it is possible and easier to discuss contraction with regard to Lie algebra. From this point of view, group contraction is defined as follows:

{\bf definition:} {\it In\"{o}n\"{u}-Wigner Contraction (Hermann \cite{Hermann}):
Given a Lie algebra $L$ associated with $G$, and a basis $X_a(a=1,2,..,n)$ of the vector space
of $L$, satisfying
$$
[X_a,X_b]=\sum_{c=1}^n D_{ab}^c X_c\qquad a,b=1,...,n.
$$
where $D_{ab}^c$ are the structure constants of $L$ with respect to this basis, the
Jacobi identity imposes the following condition on the structure constants
\begin{equation}
\sum_{c=1}^n (D_{bd}^c D_{ac}^e + D_{da}^c D_{bc}^e + D_{ab}^c D_{dc}^e)=0.
\label{form2}
\end {equation} 
For the basis of $L$, we suppose that with a continuous, directed, contraction parameter $\lambda$ 
\begin {itemize}
\item the infinite sequence ${[X_a]^\lambda}$ and the corresponding structure constants  
${[D_{ab}^c]^\lambda}$ are known,
\item the $\lim_{\lambda \rightarrow \infty}\ [D_{ab}^c]^\lambda=\ [D_{ab}^c]^{\infty}$ exists for all ${a,b,c}$, and
\item eqn.(\ref {form2}) is consistent under the limit.
\end {itemize} 
Then as a consequence ${[D_{ab}^c]^{\infty}}$ generates a new algebra $L'$ called the
contraction of $L$. 
}

With this as a starting point we investigate in the following sections the consequences of the group contraction method for representation spaces of various well-known spacetimes symmetry groups, including the de Sitter groups contracted to the Poincar\'e group.

\section {Contraction of $SO_3 \to E_2$} 

\subsection {The Algebras:}\label{so3alg}  

With the main purpose of establishing our procedure we begin with the algebra contraction $so_3
\to e_2$. The rotation group $SO_3$ in $R^3$ has three generators $\{ J_1,J_2,J_3 \}$
corresponding to infinitesimal rotations and satisfying the associated algebra
\begin{equation}
\left[ J_r,J_s \right] = i\epsilon_{rs}^t J_t \qquad r,s,t=1,2,3,
\label{alg-so3}
\end {equation} 
where $\epsilon_{rs}^t $ is an rank-$(2,1)$ tensor antisymmetric in the lower indices with $\epsilon_{12}^3=1$. To initiate the contraction, we introduce a positive real parameter $R$ and construct the following sequence of elements defined on the $so_3$ algebra, with $J_{s,R}\equiv J_{s}(R)$ (See Appendix). In principle we don't  assume the linear relation $J=R\wedge P$ of classical mechanics because it actually is a consequence of the following contraction process.
\begin{equation} 
J_0 \equiv J_3, \qquad\Pi_s \equiv {J_{s,R}\over R} \qquad
K_s \equiv \lim_{R \to \infty} \Pi_s\qquad s=1,2,
\label{susti-so3}
\end {equation} 
for which the resulting algebra  takes the form 
\begin{equation}
\left[ \Pi_1,\Pi_2 \right] = {i \over R^2} J_0, \qquad 
\left[ \Pi_2,J_0 \right] = i \Pi_1, \qquad
\left[ J_0,\Pi_1 \right] = i \Pi_2.
\label{alg-so}
\end {equation} 
On taking the limit $R\to\infty$ in eqn.(\ref{alg-so}), we obtain the algebra
$$
\left[ K_1,K_2 \right] = 0, \qquad
\left[ K_2,J_0 \right] = iK_1, \qquad 
\left[ J_0,K_1 \right] = iK_2. 
$$
These relations define the Euclidean algebra of `translations' and rotations, denoted as
$e_2\equiv t_2 \oplus so_2$. Here the generator $J_3$ associated with rotations about the
z-axis remains unchanged, while the generators $J_1$, $J_2$, determining rotations about the
$x$ and $y$ axes respectively, are transformed under the limit into generators of `translation' in
directions $y$, $x$. Interpreting this one requires the symmetry $SO_3$ about the centre
$P_{N/S}=(0,0,\pm R)$ of the sphere of radius $R$ is locally seen as an approximate symmetry
$T_2\rtimes SO_2$ of the tangent plane at the pole or the plane at infinity when $R\to\infty$
(see \cite{Izmest} for a generalization to [$O_{n+1}\to E_n$]).

Taking the Casimir of $SO_3$
$$
J^r J_r = J_1^2+J_2^2+J_3^2 = J.J,
$$
and applying eqn.(\ref {susti-so3}), we obtain  
$$
J_R.J_R=R^2(\Pi_1^2+\Pi_2^2)+J_0^2.
$$
On taking the contraction limit define the Casimir for $E_2$ as
$$
C_{E_2}\equiv \lim_{R \to \infty}{J_R.J_R \over R^2}
=\lim_{R\to\infty}(\Pi_1^2+\Pi_2^2+J_0^2/R^2)=K_1^2+K_2^2=K.K.
$$
Physically these Casimirs reflect the conservation laws of angular momentum and linear momentum,
the one being the limit of the other under contraction.

In the standard treatment of $SO_3$ it is usual to transform to the spherical tensor basis of
the $so_3$ algebra
$$
J_+\equiv J_1+iJ_2,\qquad J_-\equiv J_1-iJ_2,\qquad J_{\pm}{}^\dagger=J_{\mp},
$$
where the generators satisfy the algebra
$$
\left[ J_3,J_+\right] = +J_+, \qquad
\left[ J_+,J_- \right] = 2J_3, \qquad 
\left[ J_3,J_- \right] = -J_-, 
$$
and the Casimir is now expressed as
$$
J.J={1 \over 2}(J_+J_-+J_-J_+)+J_3^2.
$$
Applying the contraction to $J_\pm$, we obtain $K_\pm$ in $E_2$ defined as  
$$
K_+ \equiv K_1+iK_2,\qquad K_- \equiv K_1-iK_2,
\qquad K_{\pm}{}^\dagger=K_{\mp},\qquad 
K_{\pm}=\lim_{R \to \infty}{J_{\pm,R}\over R},
$$
with commutators
\begin {equation}
\left[K_+,K_-\right] = 0, \qquad 
\left[J_0,K_-\right] = -K_-, \qquad 
\left[J_0,K_+\right] = +K_+, 
\label{250}
\end {equation} where the associated Casimir takes the form
\begin {equation} K.K =K_+K_-={1 \over 2}\left \{ K_+,K_- \right \} =K_-K_+.
\label{260}
\end {equation} 

\subsection {The Representation Spaces:}\label {so3rep}  

We now investigate the effect of contraction on the representations space of $SO_3$. The bases
of the representation spaces, labelled by half-integers $j$ and $m$, are spanned by the
eigenvectors $|jm\rangle$ for $J.J$ and $J_3$ (see \cite{simetries})
\begin {eqnarray} 
J.J|jm\rangle&=&|jm\rangle  j(j+1),\qquad  J_3|jm\rangle=|jm\rangle m,\qquad \mbox{ with} \qquad -j\leq m \leq j \label{new-val}\nonumber\\ 
J_+|jm\rangle &=&|j,m+1\rangle a_{jm},\qquad \|a_{jm}\|^2=j(j+1)-m(m+1)\nonumber\\ 
J_-|jm\rangle &=&|j,m-1\rangle b_{jm},\qquad  \|b_{jm}\|^2=j(j+1)-m(m-1).\nonumber
\end {eqnarray} 
The eigenvectors and corresponding eigenvalues of $K.K$, $J_0$ in $E_2$ are given as
\begin {equation} 
K.K |km\rangle=|km\rangle k^2, \qquad\qquad J_0|km\rangle=|km\rangle m. 
\label {270}
\end {equation}  
From eqns.(\ref{270}) and (\ref{250}) we require
\begin {eqnarray} 
J_0 K_+|km\rangle&=&K_+|km\rangle(m+1)\qquad\Rightarrow 
K_+ |km\rangle = | k,m+1\rangle \bar a_{jm} \label {290}\\ 
J_0 K_-|km\rangle&=&K_-|km\rangle(m-1)\qquad\Rightarrow 
K_- |km\rangle =| k,m-1\rangle  \bar b_{jm} \label {300}
\end {eqnarray} 
Evaluating $\bar a_{jm}, \bar b_{jm}$ using eqns.(\ref{260},\ref{270}) and (\ref{290},\ref{300}) we obtain 
$$
K.K|km\rangle = K_+K_-|km\rangle =|km\rangle\bar b_{jm}\bar a_{jm} = |km\rangle k^2
\qquad\Rightarrow\quad\bar b_{jm}\bar a_{jm} = k^2.
$$
In addition using the adjoint property of the operators 
$$
\langle km | K_-K_+ |km\rangle=\langle km|km\rangle\bar a_{jm}^*\bar a_{jm} \geq 0 
\qquad\Rightarrow k^2=\bar a_{jm}^*\bar a_{jm}>0
$$
From this we deduce that
$$
\bar b_{jm}=\bar a_{jm}^* \qquad |\bar a_{jm}|=k>0.
$$

In order to establish a connection under group contraction between the two groups we need to
correlate their respective group actions. Since $J_0$ is unchanged by contraction its
group action also remains unchanged. The Casimirs provide a different situation and therefore 
the eigenvalues $j$, and hence eigenvectors, are dependent on $R$ in such a way
that $\lim_{R \to \infty}{j\over R}=k$ (see \cite{Cattaneo,Weimar,Dooley}). Thus we introduce
$j_R$ and eigenvectors $|j_Rm\rangle$ with action
\begin {eqnarray} 
\Pi_+|j_Rm\rangle&=&|j_R,m+1\rangle a_{j_Rm} \qquad ||a_{j_Rm}||^2\equiv j_R (j_R+1)-m(m+1)\nonumber\\
\Pi_-|j_Rm\rangle&=&|j_R,m-1\rangle b_{j_Rm} \qquad ||b_{j_Rm}||^2\equiv j_R (j_R+1)-m(m-1),\nonumber
\label{350}
\end{eqnarray} 
and satisfying the conditions
\begin {equation}
\lim_{R \to \infty}{j_R\over R}=k\qquad \lim_{R \to \infty}|j_Rm\rangle =|km\rangle.
\label{360}
\end {equation} 
Therefore we obtain that
\begin {eqnarray} 
K.K|km\rangle&=&\lim_{R\to\infty}{J_R.J_R\over R^2}|j_Rm\rangle \nonumber\\
\Rightarrow\ |km\rangle k^2&=&\lim_{R\to\infty}|j_Rm\rangle {j_R(j_R+1)\over R^2},\nonumber
\end {eqnarray}
We then see that from eqns.(\ref{260}) and (\ref{350})
\begin {eqnarray} 
K_+K_-|km\rangle
&=&\lim_{R\to\infty}{1\over2}\{\Pi_+,\Pi_-\}|j_R m\rangle\nonumber\\ 
&=&\lim_{R \to \infty}|j_R m\rangle{j_R (j_R+1)-m(m-1) \over R^2}\nonumber\\
&=&|km\rangle k^2.\nonumber
\end {eqnarray}
as required. From eqns.(\ref{susti-so3}) and (\ref{360}) we can infer that $j_R\sim Rk \Rightarrow \hbar j_R\sim R(\hbar k)=Rp$, in analogy to the standard classical angular
momentum definition $\vec{\j}=\vec{r}\times \vec{p}$. Note that as $R$ increases, $j_R$ takes large values
corresponding to the regime of classical physics.

\section {Contraction of $SO_{3,1} \to {\bf G}_3$} 

\subsection {The Algebras:}  

As our second application let us consider the homogeneous Lorentz group $SO_{3,1}$, for which we
have six generators $J_r$, $B_r$ $(r=1,2,3)$ obeying the following algebra
\begin {equation}
\left[B_r,B_s \right] =-i\epsilon_{rs}^t J_t, \qquad 
\left[J_r,B_s \right]=i\epsilon_{rs}^t B_t,\qquad
\left[J_r,J_s \right]=i\epsilon_{rs}^t J_t, 
\label{xyz}
\end {equation}
where $B_r$ represents the Lorentz boosts, and $J_r$ spatial rotations which clearly form a
subgroup of $SO_{3,1}$. The Casimirs of the proper Lorentz group are found to be
\begin{equation} 
C_2^{1,3}=J^rJ_r-B^rB_r=J.J-B.B,\qquad  C_4^{1,3}=J^rB_r+B^rJ_r=2J.B. 
\label{cas-lor}\\
\end{equation} 
In order to perform the contraction in eqn.(\ref{xyz}), we define the following sequence with $B_{r,c}\equiv B_r(c)$
\begin {equation} 
J_r,\qquad\Omega_r\equiv{B_{r,c}\over ic},\qquad 
G_r\equiv\lim_{c\to\infty} \Omega_r,\qquad r=1,2,3,
\label{v}
\end {equation}
where we introduce the velocity of light $c$ to parametrize the contraction procedure. 
For convenience the imaginary unit appears solely to obtain  positive valued Galilei
group Casimirs defined below. In taking the limit $c\rightarrow\infty$, the relations
given by eqn.(\ref{xyz}) take the form
\begin {equation}
\left[ G_r,G_s \right] = 0,\qquad
\left[ J_r,G_s \right] = i\epsilon_{rs}^t G_t,\qquad
\left[ J_r,J_s \right] = i\epsilon_{rs}^t J_t.\label{n}
\end {equation} 
Obviously these relations define locally the homogeneous Galilei group 
${\bf G}_3\equiv K_3\rtimes SO_3$, which contains the transformations associated with spatial
rotations $J_r$ and a change of inertial reference system $G_r$. Clearly this group is
isomorphic with the Euclidean group in 3-Dim $E_3\equiv T_3\rtimes SO_3$ but its physical
content is distinct --- one is the symmetry of 3-space and the other the symmetry group of the
laws of Newtonian motion. We find on using the substitutions given by eqn.(\ref{v}) in the
expressions for the Lorentz Casimirs given by eqn.(\ref{cas-lor}), redefining the Casimirs and
then taking the limit
\begin{eqnarray} 
C_2^3 \equiv\lim_{c\to\infty}{C_{2,\ c}^{1,3}\over c^2}
&=&\lim_{c\to\infty}(J^rJ_r/c^2+\Omega^r\Omega_r)=G^rG_r=G.G,\label{cas-gal}\\ 
\lim_{c\to\infty}{C_{4,\ c}^{1,3}\over ic}
&=&\lim_{c\to\infty}2(J^r\Omega_r)=2(J^rG_r)=2J.G.
\label{cas-gal-1}
\end{eqnarray} 
But while the former remains a Casimirs for the Galilei group, the latter is not as
$[J,J.G]\neq0$. Naturally, in the nonrelativistic limit $c\to \infty$ (or physically small
velocities $v\to 0$) the homogeneous Lorentz group is transformed into the homogeneous Galilei
group.

The usual analysis of the Lorentz group employs the isomorphism between its algebra and that of
$\overline{SL}_2\times SL_2$ where the overscore signifies that the two $SL_2$ are strictly
related by complex conjugation. We redefine the generators of the Lorentz group as
$$ 
{\cal A}_r^- \equiv {1\over2}(J_r-iB_r), \qquad{\cal A}_r^+\equiv {1\over2}(J_r+iB_r),
$$ 
with ${\cal A}_r^\pm{}^{\dagger}={\cal A}_r^{\pm}, J_r^\dagger=J_r$ and
$B_r^\dagger=-B_r$, and we obtain 
\begin {equation}
\left[ {\cal A}_r^-,{\cal A}_s^-\right] = i\epsilon_{rs}^w {\cal A}_t^-, \qquad
\left[ {\cal A}_r^+,{\cal A}_s^-\right] = 0, \qquad
\left[ {\cal A}_r^+,{\cal A}_s^+\right] = i\epsilon_{rs}^t {\cal A}_t^+. 
\label{lo-new-alg-so-c}
\end {equation} 
The Casimirs of each $sl_2$ algebra are now
$$ 
{\cal A}^-.{\cal A}^-={\cal A}^{-\ r}{\cal A}_r^-,\qquad  
{\cal A}^+.{\cal A}^+={\cal A}^{+\ r}{\cal A}_r^+.
$$ 
while the Lorentz Casimirs can be expressed as
$$ 
C_2^{1,3}=2({\cal A}^- .{\cal A}^-+{\cal A}^+.{\cal A}^+),\qquad
C_4^{1,3}=2i({\cal A}^-.{\cal A}^--{\cal A}^+.{\cal A}^+).
$$ 

As the operators $\{{\cal A}_r^-\}$ and $\{{\cal A}_r^+\}$ form two sets of independent
generators of $SL_2$ (see \cite{Ayala-Haase}), and therefore can be separately
redefined in terms of spherical vector operators ${\cal A}^+_\mu$ and ${\cal A}^-_\mu$ with
$\mu=(+,0,-)$, as in section (\ref{so3rep})
$$ 
{\cal A}^\pm_+\equiv {\cal A}_1^\pm + i {\cal A}_2^\pm,\qquad
{\cal A}^\pm_-\equiv{\cal A}_1^\pm- i {\cal A}_2^\pm,\qquad
{\cal A}^\pm_0\equiv{\cal A}_3^\pm.
$$ 

\subsection {The Finite Representations:}  

Before developing the representation theory of the Lorentz and Galilei groups, we can rewrite
the Lorentz generators as ${\cal A}_r^-\equiv \bar{\cal A}_r \otimes I$ and 
${\cal A}_r^+\equiv I\otimes {\cal A}_r$ to emphasis the fact that we are dealing with two
commuting $SL_2$ algebras. These generators act on the direct product space 
${\cal V}_{j_1}\otimes{\cal V}_{j_2}$ and correspond to two independent vector spaces under
transformations of Lorentz group. Since $SL_2\sim SO_3$ locally, we can use the results of the
previous section, and as a consequence the representations of the Lorentz group can be labeled
as $[j_1 \otimes j_2]$ with dimension $(2j_1+1)(2j_2+1)$.\\

The two $SL_2$ Casimirs ${\cal A}^\mp.{\cal A}^\mp$ have eigenspectra $j_1(j_1+1)$ and
$j_2(j_2+1)$ respectively, which establishes the eigenspectra of the two Lorentz Casimirs as
\begin{equation} 
C_2^{1,3} : 2[j_1(j_1+1)+j_2(j_2+1)]=2(j_1+j_2+1)(j_1+j_2)-4j_1j_2, \label {loA-150}
\end{equation}
\begin{equation}
C_4^{1,3} : 2i[j_1(j_1+1)-j_2(j_2+1)]=2i(j_1+j_2+1)(j_1-j_2). \label{loB-150}
\end {equation} 
The basis of the space ${\cal V}_{j_1}\otimes{\cal V}_{j_2}$ is constructed as a direct product
of vector bases of $SO_3$ with vectors $|j_1m_1,j_2m_2\rangle$. The eigenspectra of $\bar A_0$
and $A_0$ are $m_1$ and $m_2$ respectively, while we also get
\begin {eqnarray} 
\bar A_{\pm}|j_1m_1,j_2m_2\rangle
&=&|j_1m_1,j_2m_2\pm 1\rangle \sqrt{j_1(j_1+1)-m_1(m_1\pm1)},\nonumber\\ 
A_{\pm}|j_1m_1,j_2m_2\rangle
&=&|j_1m_1,j_2m_2\pm 1\rangle \sqrt{j_2(j_2+1)-m_2(m_2\pm1)}.\nonumber
\end{eqnarray}

To determine the action of $J_r$ and $B_r$ on $|j_1m_1,j_2m_2\rangle$ we follow the
development for the $so_3$ algebra in section (\ref{so3rep}). We have
$$ 
J_r={\cal A}_r^-+{\cal A}_r^+ \qquad B_r=i({\cal A}_r^--{\cal A}_r^+).
$$ 
and therefore
\begin {eqnarray} 
J_r|j_1m_1,j_2m_2\rangle
&=&{\cal A}_r^-|j_1m_1,j_2m_2\rangle+{\cal A}_r^+ |j_1m_1,j_2m_2\rangle,\nonumber\\  
B_r|j_1m_1,j_2m_2\rangle
&=&i({\cal A}_r^-|j_1m_1,j_2m_2\rangle-{\cal A}_r^+|j_1m_1,j_2m_2\rangle).\nonumber
\end {eqnarray} 
In particular for $J_3$ y $B_3$, we have
\begin {eqnarray} 
J_3|j_1m_1,j_2m_2\rangle&=&|j_1m_1,j_2m_2\rangle(m_1+m_2),\nonumber\\   
B_3|j_1m_1,j_2m_2\rangle&=&|j_1m_1,j_2m_2\rangle i(m_1-m_2).
\label{j3b3}
\end {eqnarray}
Exploiting the various isomorphisms, all the analysis for $SO_{3,1}$ is similar to the above
analysis for $SO_3$.

A separate analysis of ${\bf G}_3$, which allows us to construct an eigenbasis
appropriate to $G.G$, $J_3$ and $G_3$
\begin {eqnarray} 
G_3|g,m,s\rangle=|g,m,s\rangle s, \ \ &&\ \ J_3|g,m,s\rangle=|g,m,s\rangle m,\nonumber\\
G.G |g,m,s\rangle&=&|g,m,s\rangle g^2.
\label {lo-270}
\end {eqnarray}
The action of the other generators can be obtained using the commutator relations but as we do
not need them here we shall omit them.

With these results we now establish the connection between the two groups under the group
contraction $SO_{3,1}\to G_3$. In analogy to $SO_3$ we begin with the eigenvalues of the
respective Casimirs and we introduce eigenvalues $j_c$ and $m_c$ with eigenvectors 
$|j_{1,c}m_{1,c},j_{2,c}m_{2,c}\rangle$. Using eqns.(\ref{cas-gal}) and (\ref{loA-150}, \ref{lo-270}) 
$$
G.G |g,m,s\rangle=\lim_{c\to\infty}{J.J-B_{c}.B_{c}\over c^2} |j_{1,c}m_{1,c},j_{2,c}m_{2,c}\rangle
$$
which yields
$$ 
|g,m,s\rangle g^2 =2\lim_{c\to\infty}|j_{1,c}m_{1,c},j_{2,c}m_{2,c}\rangle {j_{1,c}^2+j_{2,c}^2+j_{1,c}+j_{2,c}\over c^2}.
$$ 
However using the quartic Casimir of eqns.(\ref{cas-gal-1}) and (\ref{loB-150}) we find
\begin{eqnarray} 
2J.G|g,m,s\rangle
&=&\lim_{c\to\infty}{2J.B_c\over ic}|j_{1,c}m_{1,c},j_{2,c}m_{2,c}\rangle\nonumber\\
&=&\lim_{c\to\infty}|j_{1,c}m_{1,c},j_{2,c}m_{2,c}\rangle{2i(j_{1,c}+j_{2,c}+1)(j_{1,c}-j_{2,c})\over ic}\nonumber\\
&=&\lim_{c\to\infty}|j_{1,c}m_{1,c},j_{2,c}m_{2,c}\rangle 2c{j_{1,c}^2+j_{2,c}^2+j_{1,c}+j_{2,c}\over c^2}\nonumber\\
&&-\lim_{c\to\infty}|j_{1,c}m_{1,c},j_{2,c}m_{2,c}\rangle 4c{j_{2,c}^2+j_{2,c}\over c^2}\nonumber
\end{eqnarray} 
Here we have rewritten the quartic Casimir eigenvalue as
$$
{2i(j_{1,c}+j_{2,c}+1)(j_{1,c}-j_{2,c})\over ic}
=2c\left ({j_{1,c}^2+j_{2,c}^2+j_{1,c}+j_{2,c}\over c^2}-2{j_{2,c}^2+j_{2,c}\over c^2}\right)
$$
for which the first term on the right yields in the limit the eigenvalue of the quadratic
Galilei Casimir $g^2$. To avoid the appearance of infinities in the limit we must then
have
\begin {equation}
\lim_{c \to \infty}{j_{1,c}\over c}=\lim_{c \to \infty}{j_{2,c}\over c}={g \over 2}.
\label {lo-400}
\end {equation}
The limiting process for the quadratic Casimir eigenvalues takes form
$$ 
g^2 = 2\lim_{c\to\infty}\left[{j_{1,c}^2+j_{2,c}^2+j_{1,c}+j_{2,c}\over c^2}\right ],
$$ 
while that of the quartic Casimir of eqn.(\ref{cas-gal-1}) and (\ref{loB-150}) we obtain
$$
\lim_{c\to\infty}{2(j_{1,c}+j_{2,c}+1)(j_{1,c}-j_{2,c})\over c}=0.
$$
and as a consequence $2J.G|g,m,s\rangle=0$. Similarly we obtain
$$
\lim_{c \to \infty} {m_{1,c}+m_{2,c}}= m, \qquad\lim_{c \to \infty}{m_{1,c}-m_{2,c}\over c}=s,
$$
and $$
|g,m,s\rangle=\lim_{c\to\infty}|j_{1,c}m_{1,c},j_{2,c}m_{2,c}\rangle
$$
by observing that under the contraction limit
$$
J_3|g,m,s\rangle=\lim_{c\to\infty}J_3|j_{1,c}m_{1,c},j_{2,c}m_{2,c}\rangle,
\quad
G_3|g,m,s\rangle=\lim_{c\to\infty}{B_{3,c}\over ic} |j_{1,c}m_{1,c},j_{2,c}m_{2,c}\rangle
$$
and hence we have
\begin {eqnarray}
|g,m,s\rangle m&=&\lim_{c\to\infty}|j_{1,c}m_{1,c},j_{2,c}m_{2,c}\rangle(m_{1,c}+m_{2,c})\nonumber \\
|g,m,s\rangle s&=&\lim_{c\to\infty}|j_{1,c}m_{1,c},j_{2,c}m_{2,c}\rangle {m_{1,c}-m_{2,c}\over c}.\nonumber
\end {eqnarray}
Introducing the linear expressions $m_{1,c}=a_1c+b_1$, $m_{2,c}=a_2c+b_2$, we find solutions
\begin{eqnarray}  
&&\lim_{c \to \infty} {(a_1+a_2)c+(b_1+b_2)}= m, \qquad
\lim_{c \to \infty} {(a_1-a_2)c+(b_1-b_2)\over c}=s,\nonumber\\  
&&\Rightarrow \qquad a_1=-a_2={s\over 2}\qquad b_1+b_2=m,\nonumber
\end{eqnarray} 
and this draws the connection between the labelling of the representation spaces of the Lorentz
and Galilei groups.

The action of generators $J_r$ and $\Omega_r$ on these new eigenvectors is as for the Lorentz
group action except for the substitution of $j$ for $j_c$ and $m$ for $m_c$ in all the matrix
elements.\\ 

From all of the above we can pass over without difficulty to the contraction of the
Poincar\'e group to the inhomogeneous Galilei group, ie. $P_{3,1}\equiv T_{3,1}\rtimes
SO_{3,1}\to G_{3,1}\equiv T_3\rtimes G_3$, given the fact that the contraction only affects the generators of Lorentz boosts, and thus contracts $SO_{3,1}\to G_3$ which has just been shown.

\section{$dS/AdS$ groups and Their Contraction to $P_{3,1}$}
\label{sittertopoincare} 

\subsection{The Algebras}\label{algebra-so32} 
The two $SO_{5-h,h}$ groups with $h=1$ or $2$ correspond to maximal group of symmetries in
De Sitter/Anti-de Sitter ($dS$/$AdS$) spaces respectively, and represent uniform and curved  space-time manifolds
that are compatible with an expanding universe. The universe can be described as a
hypersurface inside a $5$-dimensional spacetime of signature $(4,1)$ or $(3,2)$. If we use
coordinates $y^i$ with $i= 1,..,5$, the hypersurface with curvature $a$ is defined as 
$$ 
(y^1)^2+(y^2)^2 +(y^3)^2 \pm (y^4)^2 - (y^5)^2 =\eta_{ij} \,y^i y^j=-a^{-2}\qquad
\mbox{with} \quad a\equiv {1\over R}.
$$
This hypersurface is invariant under linear transformations that preserve the
metric $\eta_{ij} =diag(+++\pm-)$. These transformations comprise the Sitter/Anti-de
Sitter groups ($SO_{5-h,h}$) whose algebra consists of ${1\over 2}5(5-1)=10$ generators,
$J_{ij}$ representing generalized rotations in $E_5$. These generators
satisfy the algebra
\begin{equation}
\left[ J_{ij} ,J_{kl} \right] = i(J_{ik}  \eta_ {jl}-J_{il}  \eta_ {jk}+J_{jl}  \eta_
{ik}-J_{jk}  \eta_ {il}),
\label{sitter}
\end {equation} 
with $i,j,k,l,m=1,..,5$.\\

The two Casimirs of the de Sitter/Anti-de Sitter groups are 
\begin {equation} 
C_2 \equiv {1\over 2}J_{ij} J^{ij}, \qquad C_4 \equiv W^iW_i,
\label{cas-sitter}
\end {equation}
where $W_i$ represent a 5-dimensional vector defined as 
\begin {equation} 
W_i \equiv {1\over 2}\epsilon_{ijklm} J^{jk}J^{lm},\qquad J^{ij}=J_{kl}\eta^{ik}\eta^{jl}.
\label{pauli}
\end {equation}
$\epsilon$ is tensor totally antisymmetric with 5 index.\\

The contraction is defined by
\begin {equation}  
\Pi_{\mu}\equiv {1\over R} J_{z\mu,R},\qquad \lim_{R \rightarrow \infty} \Pi_{\mu}= K_{\mu},
\qquad\mbox {with $z=1$ or $5$}. \label{7}
\end {equation} 
with $J_{z\mu,R}\equiv J_{z\mu}(R)$.

Using eqn.(\ref {7}) we reexpress eqn.(\ref {sitter}) in a 4-dimension spacetime notation as
\begin {equation}
\left[ J_{\mu \nu} ,J_{\rho \sigma} \right] = i(J_{\mu \rho}  \eta_ {\nu \sigma}-J_{\mu \sigma}  \eta_
{\nu \rho}+J_{\nu \sigma}  \eta_ {\mu \rho}-J_{\nu \rho}  \eta_ {\mu \sigma}),
\label{a}
\end {equation}
\begin {equation}
\left[ \Pi_{\mu} ,J_{\rho \sigma} \right] = i(\Pi_{\rho}  \eta_ {\mu \sigma}-\Pi_{\sigma}  \eta_ {\mu
\rho}),\qquad \left[ \Pi_{\mu} ,\Pi_{\nu} \right] = {i \over R^2} J_{\mu \nu}, 
\label{b}
\end {equation}
with $\mu,\nu,\rho,\sigma =1,2,3,4$ (see also \cite{Gradechi}). Using eqn.(\ref{7}) in
eqns.(\ref{a},\ref{b}) we obtain the following algebra
\begin {equation}
\left[ J_{\mu \nu} ,J_{\rho \sigma} \right] = i(J_{\mu \rho}  \eta_ {\nu \sigma}-J_{\mu \sigma}  \eta_
{\nu \rho}+J_{\nu \sigma}  \eta_ {\mu \rho}-J_{\nu \rho}  \eta_ {\mu \sigma}), \label{a-1}
\end {equation}
\begin {equation}
\left[ K_{\mu} ,J_{\rho \sigma} \right] = i(K_{\rho}  \eta_ {\mu \sigma}-K_{\sigma}  \eta_ {\mu \rho}),
\qquad \left[ K_{\mu} ,K_{\nu} \right] = 0.
\label{b-1}
\end {equation}
$K_{\mu}$ denotes the translation operator in flat space-time, where a
rotation on either the surface $(x_1,x_{\mu})$ or $(x_{\mu},x_5)$ transforms as a
space-time translation in the limit of curvature zero ($R\to\infty$). The
generators $K_{\mu}$, $J_{\nu \sigma}$ along with eqns.(\ref{a-1},\ref{b-1}) define the
Poincar\'e group in the contraction. Rewriting the `de Sitter' Casimir invariants  in
(\ref{cas-sitter}) using eqn.(\ref{7}) we have
\begin {equation} 
C_2 ={1\over 2}J_{ij} J^{ij}\qquad \to \qquad C_{2,R}=R^2 \Pi_{\mu}\Pi^{\mu}+{1\over 2}J_{\mu \nu} J^{\mu \nu}.
\label{10}
\end {equation}

We define the first Casimir invariant of Poincar\'e in the limit $R\to \infty$ as   
\begin {equation} 
C^{3,1}_2 \equiv \lim_{R\rightarrow \infty} {C_{2,R} \over R^2}= K_{\mu}K^{\mu}.
\label{11}
\end {equation} 
Using eqn.(\ref{7}) in eqn.(\ref{pauli}), the second `de Sitter' Casimir invariant takes the form
\begin {eqnarray} 
C_{4,R} = W_{i,R} W_{\ R}^{i} &=&{1\over 4}R^2\epsilon_{\lambda z\rho\mu\nu}\Pi^{\rho}J^{\mu\nu}
\epsilon^{\lambda z\rho'\mu'\nu'} \Pi_{\rho'}J_{\mu'\nu'}\nonumber\\ 
&&+{1\over 4} J_{\mu \nu} J^{\mu \nu} J_{\mu' \nu'} J^{\mu' \nu'}. 
\label{13}
\end {eqnarray} 
The second term of the right-hand in eqn.(\ref{13}) vanishes in the limit of zero curvature, and
therefore the second Poincar\'e invariant takes the form
\begin {equation} 
C^{3,1}_{4} \equiv \lim_{R \rightarrow \infty} {C_{4,R}\over R^2} 
= J^{\mu\nu}J_{\mu\nu}K^\rho K_\rho - J^{\nu\rho}K_{\nu}J_{\mu\rho}K^{\mu}.
\label{cas4-sitter}
\end {equation}

\subsection {$AdS$ Unitary Representations}\label{representacion-unitaria-so32} 

Following Nicolai \cite{Nicolai}, the generators $J_{ij}\in SO_{3,2}$ permit a spinorial
representation in terms of the  gamma-matrices $\langle ..|J_{ij}|..\rangle=\Gamma_{ij}$, this matrix set  is written as
\begin {eqnarray} 
\Gamma_{ij}&=&\Gamma_{ij}^\dagger \qquad \hbox{for }\ J_{rs},\ J_{45} \quad \hbox{with }\ r,s= 1,2,3, \nonumber\\ 
\Gamma_{ij}&=&-\Gamma_{ij}^\dagger \quad\hbox{for }\ J_{4r},\ J_{r5}.  
\label{nohermitico}
\end {eqnarray}
We see that not all the $J_{ij}s$ are unitary, in accordance with the fact that
finite-dimensional representations of a non-compact group  are not unitary representations. To
obtain a Hermitian representation of the generators $J_{ij}=J_{ij}^\dagger$, a 
infinite-dmensional representation is required. From the eqn.(\ref{nohermitico}) we distinguish
the compact generators $J_{45}$ and $J_{rs}$ from the non-compact generators $J_{r5}$ and
$J_{4r}$. The operators $J_{rs}$ and
$J_{45}$ generate a maximal compact subalgebra associated with $SO_3\times SO_2$.  
We now define the following operators in $SO_{3,2}$ as
\begin {equation}
M_{r}^+ \equiv i J_{4r}+ J_{5r}\ , \qquad  M_{r}^- \equiv i J_{4r}- J_{5r}, \quad \mbox{where}\quad M_{r}^-=-(M_{r}^+)^\dagger,
\label{boost}
\end {equation} 
with $r=1,2,3 $. The associated algebra is given as
\begin {eqnarray}
\left [M_{r}^+,M_{s}^- \right ]&=& 2(\delta_{rs} J_{45} + iJ_{rs}), \qquad    
\left [M_{r}^+,M_{s}^+ \right ]= \left [M_{r}^-,M_{s}^- \right ]= 0,
\nonumber\\
\left [J_{45},M_{r}^+ \right ]&=& M_{r}^+, 
\qquad
\left [ J_{45},M_{r}^- \right ]=-M_{r}^-.  
\label{conm-pm}
\end {eqnarray}
The laddering operators $M_r^+$ and $M_r^-$ respectively raise and lower the energy eigenvalues
in unit of one when applied to the eigenstates of the energy operator $J_{45}$. Choosing 
$J_{45}$, $J^rJ_r$ and $J_3$ as our set of mutually commuting operators we identify our
eigensystem as follows:
\begin {eqnarray} J^{r}J_{r}\ |(...)E_0\ s\ m\rangle
&=&s(s+1)\ |(...)E_0,\ s,\ m\rangle, \nonumber\\ 
J_{45}\ |(...)E_0\ s\ m\rangle&=&E_0\ |(...)E_0,\ s,\ m\rangle,\nonumber\\ 
J_{3}\ |(...)E_0\ s\ m\rangle&=&m\ |(...)E_0,\ s,\ m\rangle, 
\label{t-ecua-propia}
\end {eqnarray}
where $(...)$ denote a non-specified set of labels.\\

The $C_2$ Casimir can be written as:
\begin {eqnarray} 
C_2&=&(J_{45})^{2} + {1\over 2}J^{rs}J_{rs}+J^{4r}J_{4r} + J^{5r}J_{5r}\nonumber\\ 
&=&(J_{45})^{2} + J^{r}J_{r} + {3\over 2} \{M_{r}^+,M_{r}^- \}. \label{newc2}
\end {eqnarray} 
Note that $\frac {1}{2} J^{rs}J_{rs}=J^rJ_r$, $J^{4r}J_{4r}=-3(J_{4r})^2$,
$J^{5r}J_{5r}=-3(J_{5r})^2$ and $\{M_{r}^+,M_{r}^-\}=-2(J_{4r}^2+J_{5r}^2)$. Taking the
representations for which exists a state set that annihilates $M_r^-$, where the energy
spectrum has a lower bound. If we denote the lowest energy eigenvalue by $E_0$ and the angular
momentum values by $s$, the ground state consists of $(2s+1)$ states
$|(E_0,s)E_0,\ s,\ m\rangle$, $m=-s,-s+1,...,s$. To evaluate $C_2$ on the ground
state $|E_0,s\rangle$, we use
\begin {equation} 
M_{r}^-\ |(E_0,s)E_0,\ s,\ m\rangle=0.
\label{ecua-propia}
\end {equation} 
Replacing $\{M_r^+,M_r^-\}$ in eqn.(\ref{newc2}) by $-[M_r^+,M_r^-]$ and using eqns.(\ref{conm-pm},\ref{t-ecua-propia})
and (\ref{ecua-propia}), we obtain
$$ 
C_2|(E_0,s)E_0,\ s,\ m\rangle=E_0(E_0-3)+s(s+1)|(E_0,s)E_0,\ s,\ m\rangle.
$$

Using eqns.(\ref{7}, \ref{t-ecua-propia}) we introducing $E_{0,R}$ in the contraction and
obtain
\begin {eqnarray} 
K_\mu K^\mu|k,\lambda\rangle
&=&\lim_{R\rightarrow\infty}{C_{2,R}\over R^2}|(E_{0,R},s)E_{0,R},\ s,\ m\rangle \nonumber\\ 
&=&\lim_{R\rightarrow\infty}{E_{0,R}(E_{0,R}-3)\over R^2}|(E_{0,R},s)E_{0,R},\ s,\ m\rangle,
\nonumber
\end {eqnarray} 
and using $E_{0,R}=Rk\ $ with $k=mc/\hbar$ ($m\Rightarrow$ rest-mass), the contraction is given
as
$$ 
K_\mu K^\mu |k,\lambda\rangle =\lim_{R\rightarrow\infty}{E_{0,R}^2\over R^2}
|(E_{0,R},s)E_{0,R},\ s,\ m\rangle\ =\ k^2|k,\lambda\rangle. 
$$

\section*{Appendix}
Since the classic angular moment is defined as $\vec{\j}= \vec{r} \wedge \vec{p}$, an infinitesimal rotation is generated when $|\vec{r}|\to \infty$, with $\vec{\theta}$ the angle and $\vec{\ell}$ arc length, where
$$
\vec{\j}.\vec{\theta}= \vec{r} \wedge \vec{p}.{\vec{\ell}\over |\vec{r}|}= {\vec{r} \wedge \vec{p}\over |\vec{r}|}.\vec{\ell}=(\vec{e}\wedge \vec{p}).\vec{\ell} \qquad \mbox{ with } \vec{e}={\vec{r} \over |\vec{r}|} 
$$
Without loss of majority, in the poles we obtain that 
$$
{\vec{\j}_i\over |\vec{r}|}={\vec{r}\wedge\vec{p_i}\over |\vec{r}|}=\vec{k}\wedge\vec{p_i}=(-p_y\vec{i} \quad \mbox{or}\quad p_y\vec{j})
$$
In the quantization process we obtain a $R-$ dependent operator $J=R\wedge P$, with $R$ and $P$ the position and momentum operators respectively, and  where an infinitesimal rotation is rewritten as
$$
J.\vec{\theta}= R \wedge P.{\vec{x}\over |R|}= ({e \wedge P}).\vec{x}, 
$$
with $|R|$ the norm of the operator $R$. In the limit $|R|\to \infty$ we obtain `the momentum operator'. In the previous sections we defined $K_i=e\wedge\kappa_i$, with $P_i=\hbar \kappa_i$

\section*{Acknowledgments}
The firt author was partially supported by the {\it Fundaci\'on Mazda para el Arte y la Ciencia} and the {\it Universidad de los Andes}. He would like to express his gratitude to the organizers of the Summer School on Geometric and Topological Methods for Q.F.T. - 2001, for the invitation to take part in the school.


\begin{thebibliography}{99} 
\bibitem{Inonu-Wigner} In\"on\"u E., Wigner E. P., {\it Proc. Natl. Acad.
Sci.,USA} {\bf 39}-510 (1953) and {\bf 40}-119 (1954).

\bibitem{Inonu} In\"on\"u E., ``Contractions of Lie groups and their
representations'' in {\it Group theoretical concepts in elementary particle physics}, 
F.G\"ursey ed., Gordan and Breach, pp.391-402 (1964).

\bibitem{Gilmore} Gilmore R., {\it Lie Groups, Lie Algebras, and some of their
applications}, J.Wiley, New York, 1974.

\bibitem{Hermann} Hermann R., {\it Lie Groups for Physicists}, Amsterdam, W.
A. Benjamin, 1966.

\bibitem{Izmest} Izmest\'ev A. A., Pogosyan G. S., Sissikian A. N., 
``Contractions on the Lie algebras and separation of variables. The n-dimensional
sphere'', {\it J. Math. Phys.} {\bf 40} 1549-1573 (1999).

\bibitem{simetries} Greiner W., Mull\"er B., {\it Quantum Mechanics
Symmetries}, New York, Springer, 1994.

\bibitem{Cattaneo} U. Cattaneo and W. Wreszinski, ``On contraction of Lie
algebra representations'', {\it Commun. Math. Phys.} {\bf 68}, 83-90 (1979).

\bibitem{Weimar} E. Weimar-Woods, ``Contraction of Lie algebra
representations'', {\it J. Math. Phys.} {\bf 32}, 2660-2665 (1991).

\bibitem{Dooley} Dooley, A. H.; Rice, J.W., ``On contractions of semisimple Lie groups'', {\it Trans. Amer. Math. Soc.} 
{\bf 289}, no. 1, 185-202 (1985).

\bibitem{Ayala-Haase} Ayala M., Haase R., ``$OSp(N|4)$ group and their contractions to
$P(3,1)\times Gauge$'', {\bf hep-th}/0102030.

\bibitem{Gradechi} El Gradechi A. M., and De BiSvre S., ``Space Quantum
Mechanics on The Anti De Sitter Space-Time and its Poincar\'e Contraction'',
{\it Annals Phys.} 235 (1994) 1-34 ({\bf hep-th}/9210133).

\bibitem{Nicolai} Nicolai H., ``Representations of supersymmetry in anti-de
Sitter space'', {\it Supersymmetry and Supergravity '84, Proceedings of the
Trieste Spring School}. World Scientific, 1984.

\end{thebibliography}
\end{document}